\documentclass[jcp,showpacs,showkeys,twocolumn]{revtex4}

\usepackage{amsmath}
\usepackage{amsfonts}            
\usepackage{amssymb}

\usepackage{mathrsfs}

\usepackage{eufrak}

\usepackage{graphicx}
\usepackage{epstopdf}

\usepackage{dcolumn}
\usepackage{bm}

\makeatletter
\def\@dotsep{4.5}
\makeatother

\begin{document}

\title{When diffraction rules the stereodynamics of rotationally inelastic collisions}

\author{Mikhail Lemeshko}
\email{mikhail.lemeshko@gmail.com}
\affiliation{%
Fritz-Haber-Institut der Max-Planck-Gesellschaft, Faradayweg 4-6, D-14195 Berlin, Germany
}%

\author{Pablo G. Jambrina}
\email{pablojambrina@gmail.com}
\affiliation{%
Departamento de Qu\'{\i}mica F\'{\i}sica, Universidad de Salamanca, 37008 Salamanca, Spain, and \\ Departamento de Qu\'{\i}mica F\'{\i}sica, Facultad de Qu\'{\i}mica, Universidad Complutense, 28040 Madrid, Spain
}%

\author{Marcelo P. de Miranda}
\email{M.Miranda@leeds.ac.uk}
\affiliation{%
School of Chemistry, University of Leeds, Leeds LS2 9JT, United Kingdom
}%

\author{Bretislav Friedrich}

\email{brich@fhi-berlin.mpg.de}

\affiliation{%
Fritz-Haber-Institut der Max-Planck-Gesellschaft, Faradayweg 4-6, D-14195 Berlin, Germany}%

\date{\today}

\begin{abstract}

Following upon our recent work on vector correlations in the Ar--NO collisions [PCCP {\bf 12}, 1038 (2010)], we compare model results with close-coupling calculations for a range of channels and collision energies for the He--NO system. The striking agreement between the model and exact polarization moments indicates that the stereodynamics of rotationally inelastic atom-molecule collisions at thermal energies is governed by diffraction of matter waves from a two-dimensional repulsive core of the atom-molecule potential.  Furthermore, the model polarization moments characterizing the He--NO, He--O$_2$, He--OH, and He--CaH stereodynamics are found to coalesce into a single, distinctive pattern, which can serve as a ``fingerprint" to identify diffraction-driven stereodynamics in future work.

\end{abstract}

\pacs{34.10.+x, 34.50.-s, 34.50.Ez}
\keywords{Rotationally inelastic scattering, stereodynamics, vector correlations, models of molecular collisions, diffraction.}

\maketitle

The pioneering work of Herschbach and coworkers~\cite{HerschbachCorrelations} on vector correlations in the domain of molecular collisions spurred an effort to extract directional information hidden in molecular dynamics experiments and computations, and thereby to enrich our knowledge of \emph{how} a given collision proceeds~\cite{QCTquant}. However, even when characterized to the full by vector correlations, the \emph{why} of collision dynamics can only be answered as well as the theoretical method applied to treat the collision allows~\cite{VectCorrReviews}. Therefore, we implemented an analytic model of collision dynamics, capable of answering the \emph{why} for a class of collisions in detail, and used it to develop an analytic model of vector correlations in such collisions~\cite{LemFriFraunhoferB}. The collision model is based on the Fraunhofer scattering of matter waves~\cite{EarlyFraunhofer}, recently extended to treat collisions in fields~\cite{LemFriFraunhoferA} as well as the stereodynamics of rotationally inelastic atom--molecule collisions at thermal and hyperthermal energies~\cite{LemFriFraunhoferB}. The Fraunhofer model~\cite{EarlyFraunhofer, LemFriFraunhoferA, LemFriFraunhoferB} relies on the sudden approximation, which treats the rotational motion as frozen during the collision and thereby allows to replace the inelastic scattering amplitude with the elastic one. The elastic scattering amplitude itself is approximated by the amplitude for Fraunhofer diffraction of matter waves from a sharp-edged, impenetrable obstacle acting in place of the molecular scatterer and captures forward scattering. At collision energies of hundreds of cm$^{-1}$, consistent with the sudden approximation, the shape of the scatterer is approximated by the repulsive core of the atom--molecule potential, with the attractive part disregarded. The Fraunhofer model renders fully state- and energy-resolved scattering amplitudes and all the quantities that unfold from them in analytic form.

The vector correlations obtained from the Fraunhofer model were found to closely reproduce those extracted from close-coupling calculations and experiment of Wade~\textit{et al.}~\cite{Wade04} for the much examined Ar--NO ($X^2 \Pi$) system~\cite{ArNOstudies,Alexander99}. This agreement allowed interpreting the system's collision stereodynamics in terms of the Fraunhofer model. The Fraunhofer model of vector correlations henceforth revealed that the stereodynamics of the Ar--NO rotationally inelastic collisions is dominated by the diffractive part of the scattering amplitude which is governed by a single Legendre moment characterizing the anisotropy of the hard-core part of the system's potential energy surface. Given the ``geometric'' origin of this behavior -- ordained by the angular momentum disposal -- we wondered about its generality. In this work, we compare model results with close-coupling calculations for different channels and collision energies of the He--NO system. The striking agreement between the Fraunhofer model and the close coupling calculations found herein and in our previous work on the Ar--NO collision stereodynamics~\cite{LemFriFraunhoferB} attests to the dominant role of diffraction in rotationally inelastic collisions. Furthermore, the model alignment moments were found to exhibit remarkable similarities for different collision partners, such as He--NO, He--O$_2$, He--OH, and He--CaH, which allowed us to identify the form factors (``fingerprints'') of diffraction-driven stereodynamics.

We note that the discrepancy between the model and the exact calculations of the differential cross sections arises from the non-diffractive contributions to scattering rather than from the neglected diffraction by the long-range potential~\cite{LemFriFraunhoferB}.

\begin{figure*}
\includegraphics[width=14cm]{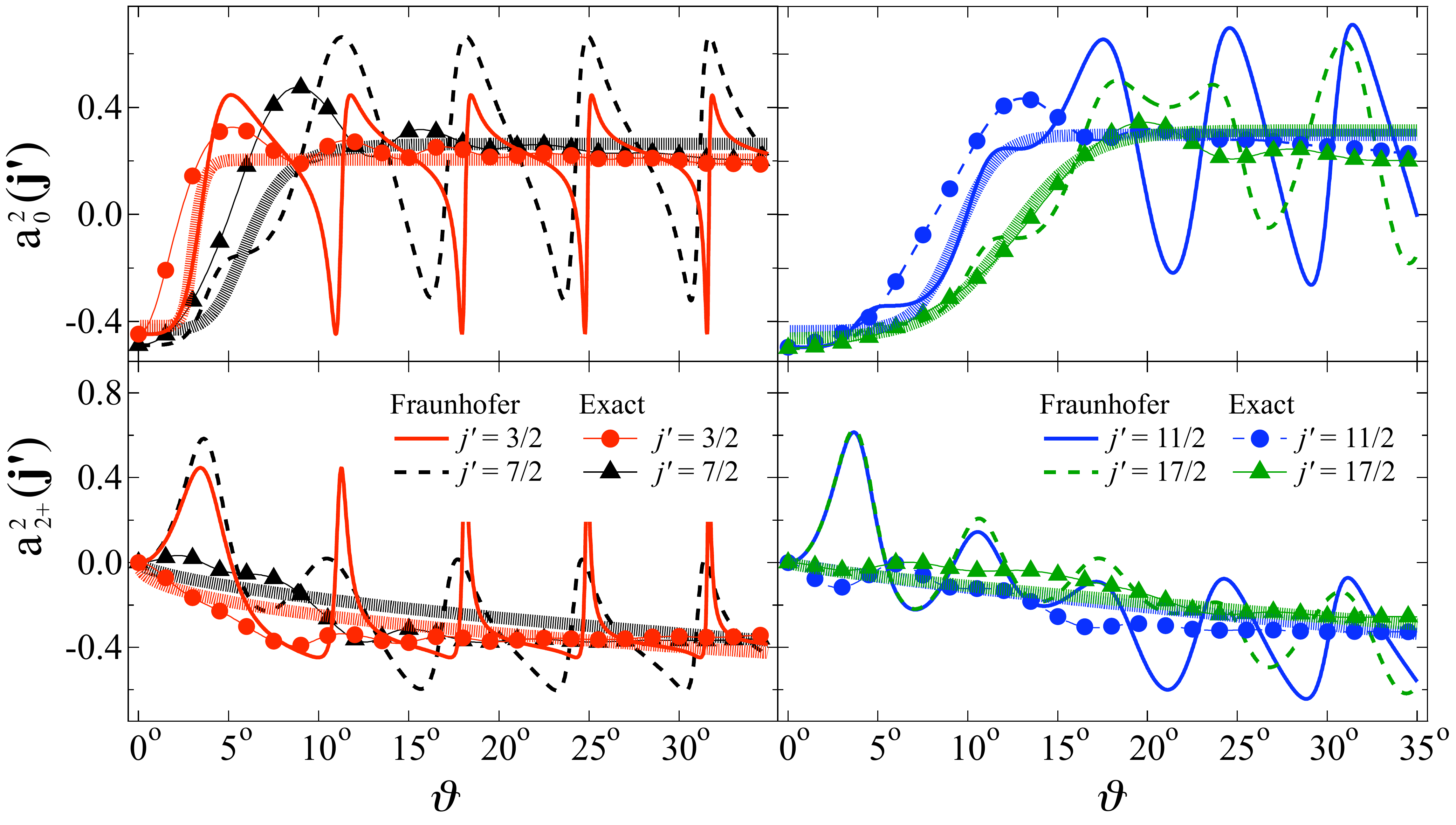}
\caption{\label{fig:channels} Polarization moments pertaining to the He--NO~$(j=\Omega=1/2 \to j', \Omega'=1/2)$ system at 520~cm$^{-1}$. Model results are shown by lines, exact computations -- by symbols. Form factors of the alignment moments are shown by thick semitransparent lines.}
\end{figure*}

The stereodynamics of an atom--diatom collision is usually described by a set of four vectors: the initial and final relative velocities, $\mathbf{k}$ and $\mathbf{k'}$, and the initial and final rotational angular momenta of the diatomic molecule, $\mathbf{j}$ and $\mathbf{j'}$.  We use the initial and final relative velocities $\mathbf{k}$ and $\mathbf{k'}$ to define the $XZ$ plane of the space-fixed coordinate system, with the initial relative velocity $\mathbf{k}$ pointing along the $Z$ axis. In keeping with the convention of de~Miranda~\textit{et al.}~\cite{MarceloMoments}, we characterize the spatial distribution of the angular momenta relative to the $XYZ$ frame by the real polarization moments $a^{k}_{q\pm}$, which are related to the multipole moments in the expansion of the density operator in terms of the irreducible tensor operators~\cite{BlumBook}. Since, within the Fraunhofer model, the scatterer is two-dimensional, the model can only account for even-$k$ (alignment) polarization moments with even $q$~\cite{LemFriFraunhoferA,LemFriFraunhoferB}. The ``extra symmetry'' of the model causes odd-$k$ (orientation) moments as well as odd-$q$ alignment moments to vanish.

First, we focus on vector correlations in the He--NO system, for different scattering channels and collision energies. We compare analytic model results with fully quantum close-coupling calculations performed using the HIBRIDON suite of computer codes~\cite{hibridon} on the PES of K\l os \textit{et al.}~\cite{HeNOPES}. In order to characterize the $\mathbf{k-k'-j'}$ three-vector correlation, we make use of the alignment moments $a^{2}_0 (\mathbf{j'})$ and $a^{2}_{2+} (\mathbf{j'})$ of the diatomic's final rotational angular momentum $\mathbf{j'}$ with respect to the $XYZ$ frame.  The physical meaning and range of the polarization moments is given in Table~\ref{table:moments}.

\begin{table}
\centering
\caption{ The physical meaning and range of the $a^{2}_0 (\mathbf{j'})$  and $a^{2}_{2+}
(\mathbf{j'})$ alignment polarization moments. The $Z$ axis points along the
initial relative velocity $\mathbf{k}$. The final relative velocity
$\mathbf{k'}$ lies in the $X>0$ half of the $XZ$ plane. The $a^{2}_0 (\mathbf{j'})$ moment accounts for alignment of $\mathbf{j'}$ with respect to $\mathbf{k}$. Positive (negative) values of $a^{2}_0 (\mathbf{j'})$ correspond to $\mathbf{j' \parallel k}$ ($\mathbf{j' \perp k}$, in which case $\mathbf{j'}$  lies in the $XY$ plane). The positive (negative) values of the $a^{2}_{2+} (\mathbf{j'})$ moment correspond to alignment of $\mathbf{j'}$ along the $X$-axis ($Y$-axis). The $a^{2}_{2-}
(\mathbf{j'})$ moment vanishes identically. The indicated
ranges of the moments correspond to the high-$j$ limit. }
\vspace{0.2cm}
\label{table:moments}
\begin{tabular}{| c | c | c | }
\hline 
\hline
Moment & $a^{2}_0 (\mathbf{j'})$  & $a^{2}_{2+} (\mathbf{j'})$ \\[3pt]
\hline
Meaning &   $\mathbf{j'}$ along $Z$ & $\mathbf{j'}$ along $Y$ or $X$ \\[0.3cm] 
Range & $\mathbf{j'} \perp Z$ \hspace{0.01cm} $\to$ \hspace{0.01cm} $-1/2$ &  \hspace{0.05cm} $\mathbf{j'} \parallel Y$ \hspace{0.01cm} $\to$ \hspace{0.01cm} $-\sqrt{3}/2$  \\[0.1cm]
  & $\mathbf{j'} \parallel Z$ \hspace{0.01cm} $\to$ \hspace{0.01cm} $1$  & $\mathbf{j'} \parallel X$  \hspace{0.01cm} $\to$ \hspace{0.01cm} $\sqrt{3}/2$   \\
 \hline
 \hline
\end{tabular}
\end{table}

Figure~\ref{fig:channels} shows the $a^2_0 (\mathbf{j'})$ and $a^2_{2+} (\mathbf{j'})$ moments for different channels of the He--NO $(j=\Omega=1/2 \to j', \Omega'=1/2)$ system at a collision energy of 520~cm$^{-1}$. All the alignment moments we present below were obtained for unresolved initial and final lambda-doublet states. The compelling agreement of the Fraunhofer model with the exact calculations for the He--NO as well as for Ar--NO~\cite{LemFriFraunhoferB} systems  attests to the predominant role of diffraction in shaping the stereodynamics of rotationally inelastic collisions at thermal and hyperthermal energies. Furthermore, the agreement with the model shows that what matters the most -- as far as the PES is concerned -- is the 2D contour of its repulsive core. Save for the moments' oscillatory structure, which differs for different final states (as these arise due to different sets of the PES's Legendre moments), the diffraction manifests itself in the same way in all the scattering channels, i.e., it leaves behind the same fingerprints, shown in Fig.~\ref{fig:channels} by the thick semi-transparent lines: whereas the $a^2_0$ moments are negative for a purely forward scattering and increase at larger scattering angles where they approach a constant positive value, the $a^2_{2+}$ moments drop from a zero for forward scattering and tend to a constant negative value at larger scattering angles. This means that, due to geometric reasons, $\mathbf{j'}$ aligns perpendicular to $\mathbf{k}$ for a purely forward scattering. At larger scattering angles, $\mathbf{j'}$ aligns along the $Z$ and $Y$ axes, indicating that the molecular axis aligns preferentially in the $X$ direction.

\begin{figure}
\includegraphics[width=7.5cm]{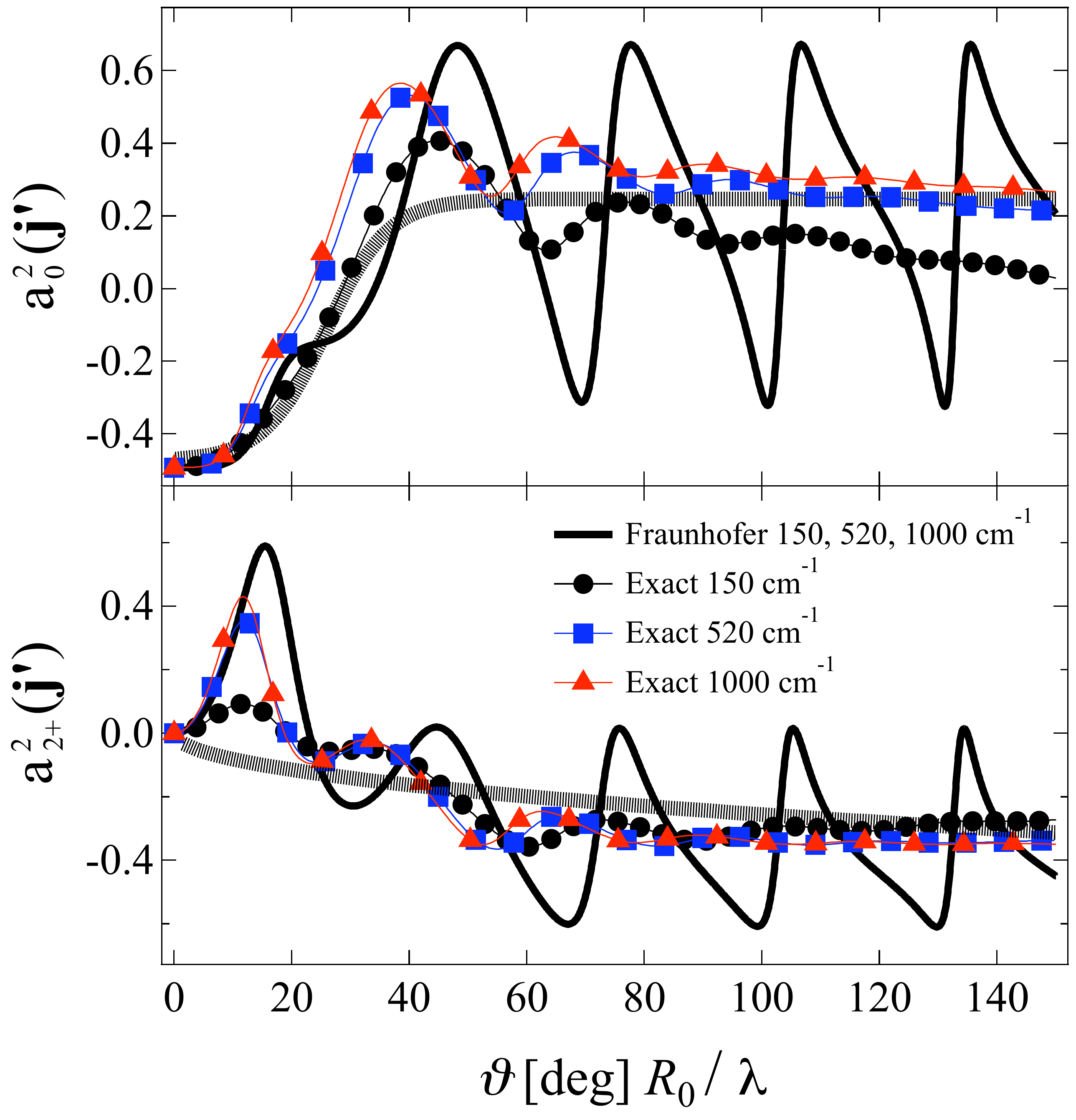}
\caption{\label{fig:energies} Polarization moments versus scaled scattering angle pertaining to the He--NO~$(j=\Omega=1/2 \to j'=9/2, \Omega'=1/2)$ system at different collision energies. The Fraunhofer moments coincide exactly. Form factors of the alignment moments are shown by thick semitransparent lines. See text.}
\end{figure}
\begin{figure}
\includegraphics[width=7.5cm]{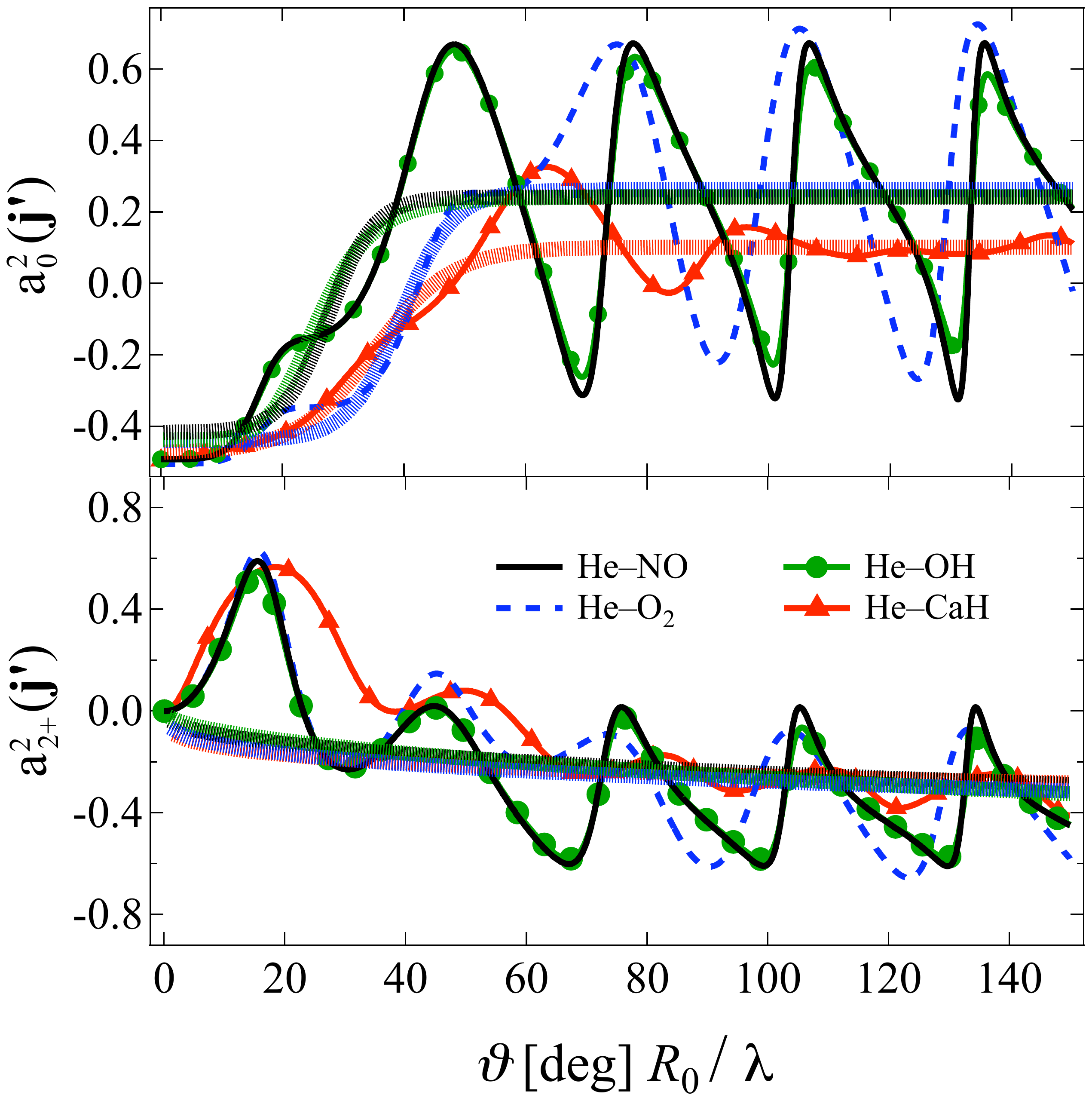}
\caption{\label{fig:molecules} Polarization moments versus scaled scattering angle pertaining  to the He--NO~$(j=\Omega=1/2 \to j'=9/2, \Omega'=1/2)$,  He--O$_2$~$(j=0, N=1, \to j'=4, N'=5)$, He--OH~$(j=\Omega=1/2, \to j'=9/2, \Omega'=1/2)$, and He--CaH~$(j=1/2, N=0, \to j'=11/2, N'=6)$ systems at 520~cm$^{-1}$. The moments were obtained using the Fraunhofer model. Form factors of the alignment moments are shown by thick semitransparent lines. See text.}
\end{figure}

If the oscillations of the alignment moments are due to diffraction, they should scale with the de Broglie wavelength of the collision system and the size of the molecular scatterer, in analogy with the wavelength of light and the obstacle size in optics. Indeed, Figure~\ref{fig:energies} reveals such a scaling of the $a^2_0 (\mathbf{j'})$ and $a^2_{2+}(\mathbf{j'})$ moments for the He--NO $(j=\Omega=1/2 \to j'=9/2, \Omega'=1/2)$ channel and a range of collision energies. The scaling was implemented by the transformation $\vartheta \rightarrow \vartheta R_0/\lambda$, where $\vartheta$ is the scattering angle, $R_0$ is the molecular size and $\lambda$ is the de Broglie wavelength. Whereas the alignment moments furnished by the Fraunhofer model coincide exactly upon scaling in $\vartheta$  (black solid line), such a scaling brings the exact moments (filled circles, squares, and triangles) quite close to one another, but does not result in their exact matching. The remaining differences among the scaled exact moments corresponding to different collision energies attest to non-diffractive contributions to scattering, the influence of the attractive branch of the PES, and a breakdown of the sudden approximation. For instance, the scaled exact moments corresponding to 520 and 1000 cm$^{-1}$ come close to one another, while the oscillations for 150 cm$^{-1}$ are quite off, especially for the $a^2_0 (\mathbf{j'})$ moment. We ascribe this discrepancy to the potential well of 25 cm$^{-1}$~\cite{HeNOPES} that brings about non-diffractive contributions to the stereodynamics while, at the same time, diminishing the role of the PES's ``repulsive core.''  However, the form factor of the alignment moments (thick semitransparent line) is quite similar to the ones shown in Figure~\ref{fig:channels}, indicating that diffraction leaves behind the same ``fingerprint'' for different collision energies and scattering channels. Interestingly, at low collision energies (10 cm$^{-1}$), the oscillations of the alignment moments fall further out of phase, but the form factors still remain in place.

In order to see how the diffraction patterns change from one scattering system to another, we also examined the He--O$_2$, He--OH, and He--CaH systems at a collision energy of 520~cm$^{-1}$ and similar channels, using potential energy surfaces of refs.~\cite{PESHe-O2,PESHe-OH,PESHe-CaH}. 

Figure~\ref{fig:molecules} shows the dependence of the polarization moments on the scaled scattering angle obtained from the  Fraunhofer model, pertaining to the He--NO~$(j=\Omega=1/2 \to j'=9/2, \Omega'=1/2)$,  He--O$_2$~$(j=0, N=1, \to j'=4, N'=5)$, He--OH~$(j=\Omega=1/2, \to j'=9/2, \Omega'=1/2)$, and He--CaH~$(j=1/2, N=0, \to j'=11/2, N'=6)$ systems. One can see that the shape of the oscillations differs for different collision partners. Going from the most symmetric scatterer, O$_2$, through NO and OH to the most asymmetric one, CaH, one can see that the oscillations become increasingly asymmetric too, and their amplitudes decrease. This suggests relating the asymmetry and amplitude of the oscillations to the asymmetry of the repulsive core of the PES. However, the form factors (shown by the semi-transparent curves) that capture the alignment moments are very similar to one another, indicating that the ``fingerprints'' of diffraction barely depend on the collision system. We note that the same can be said about the higher-rank alignment moments, which we evaluated as well~\cite{PabloToBePublished}.

In summary, in order to gain insight into the stereodynamics of rotationally inelastic atom-molecule collisions, we compared polarization moments obtained from an analytic model with those extracted from exact close-coupling calculations for the He--NO collisions. The model alignment moments were found to come as close to exact results as for the previously examined Ar--NO system~\cite{LemFriFraunhoferB}, which reveals that the collision stereodynamics in question is governed by diffraction of matter waves from a 2D contour of the repulsive core of the potential (flat sharp-edged obstacle). The oscillatory patterns of the alignment moments due to diffraction scale with the de Broglie wavelength and the molecular size. Therefore, deviations from such patterns single out other contributions to the scattering which are mainly non-diffractive. Furthermore, diffraction leaves behind the same fingerprints for different channels and collision energies for a range of systems, including He--NO, He--O$_2$, He--OH, and He--CaH. These fingerprints can be used to identify diffraction-driven stereodynamics in future experiments and exact computations.

We thank Javier Aoiz for helpful comments, Mark Brouard for stimulating discussions, and Gerard Meijer for encouragement and generous support. Financial support of the Spanish Ministry of Science and Innovation (grant CTQ2008-02578) is gratefully acknowledged.



\end{document}